\documentclass[prl, superscriptaddress, reprint, showpacs, twocolumn, balance, notitlepage]{revtex4-1}

\usepackage{graphicx} 
\usepackage[normalem]{ulem}
\usepackage{tikz}
\usepackage{amsmath}
\usepackage{amssymb}
\usepackage{float}
\usepackage{color}
\usepackage{comment}

\definecolor{blueJay}{rgb}{0.541, 0.741, 1}
\definecolor{yellowJay}{rgb}{1, 0.753, 0}
\definecolor{lightBlueJay}{rgb}{0, 0.992, 1}
\definecolor{purpJay}{rgb}{0, 0.125, 0.376}
\definecolor{lightBlue}{rgb}{0, 0.69, 0.941}
\definecolor{midnightBlue}{rgb}{0.1, 0.1, 0.44}

\begin{document}
\raggedbottom 
\title{Packing Transitions in the Elastogranular Confinement of a Slender Loop}
\author{David J. Schunter, Jr., Regina K. Czech, and Douglas P. Holmes}
\affiliation{\footnotesize Mechanical Engineering, Boston University, Boston, MA, 02215, USA} 

\date{\today}

\begin{abstract} 
Confined thin structures are ubiquitous in nature. Spatial and length constraints have led to a number of novel packing strategies at both the micro-scale, as when DNA packages inside a capsid, and the macro-scale, seen in plant root development and the arrangement of the human intestinal tract. By varying the arc length of an elastic loop injected into an array of monodisperse, soft, spherical grains at varying initial number density, we investigate the resulting packing behaviors between a growing slender structure constrained by deformable boundaries. At low initial packing fractions, the elastic loop deforms as though it were hitting a flat surface by periodically folding into the array. Above a critical packing fraction $\phi_c$, local re-orientations within the granular medium create an effectively curved surface leading to the emergence of a distinct circular packing morphology in the adjacent elastic structure. These results will bring new insight into the packing behavior of wires and thin sheets and will be relevant to modeling plant root morphogenesis, burrowing and locomotive strategies of vertebrates \& invertebrates, and developing smart, steerable needles. 
\end{abstract}
\maketitle

Under rigid confinement, thin structures tend to adopt the geometry circumscribed by their confining boundaries~\cite{stoop2008, hure2012, paulsen2019, davidovitch2019, cerda2004}. Draping a thin sheet or filament onto a rigid flat surface causes it to fold~\cite{ribe2003, sano2017}, leading to the formation of multiple alternating loops as the arclength is continuously increased~\cite{lloyd1978, mahadevan1999}. In the presence of a rigid curved surface, flexible structures may coil, roll-up, or spiral, as seen with the packaging of household paper products~\cite{romero2008} or when pulling a thin sheet through a small aperture~\cite{boue2006}. Similar folding and circular morphologies have also been observed in thin structures under flexible/soft confinement~\cite{vetter2015, shaebani2017}. In both types of confinement, the material and structural characteristics of the containing space (geometry, rigidity, etc.) are effectively fixed: even when softly confined, packing thin structures can only slightly influence their flexible containers~\cite{vetter2014}. What happens when the notions of confining rigidity and geometry are less clearly defined is not well understood, yet this situation frequently occurs when slender objects pack within complex and fragile media.

\begin{figure*}
\resizebox{1.0\textwidth}{!}{\includegraphics{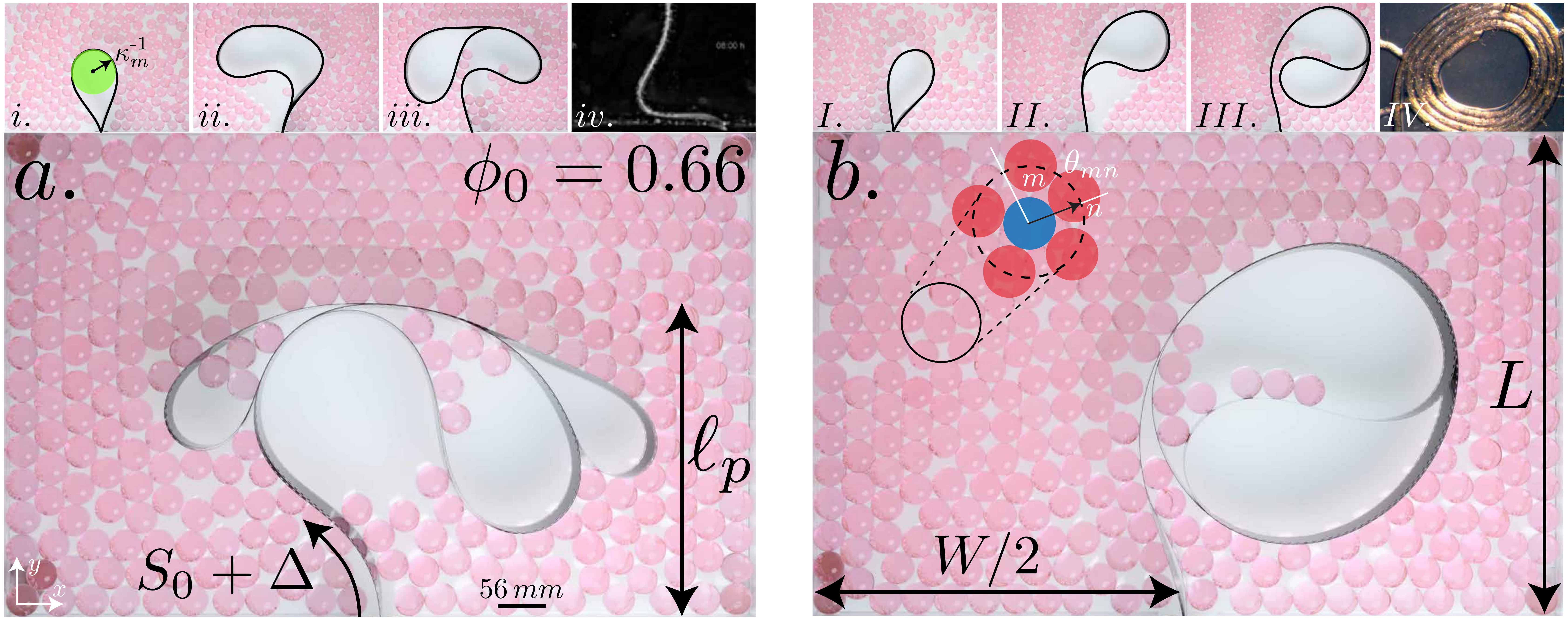}}
\caption[]{A slender elastic loop quasistatically injected into a granular array of varying initial packing fraction $\phi_0$. (a) Below a critical initial packing fraction $\left(\phi_0<0.641\right)$, the elastic loop will pack into the granular medium by adopting a characteristic folded geometry. (b) For $\phi_0\geq 0.641$, a characteristic circular packing morphology is seen to emerge. At low values of injected arclength $\Delta$, the two morphologies appear the same (i,I). Once set, the resulting circular and folded morphologies bear a striking resemblance to developing $\textit{Aribidopsis}$ roots in contact with a hard agar substrate (iv,IV) [Adapted from~\cite{kolb2017, migliaccio2001}].}
\label{fig1}
\end{figure*}

Drawing inspiration from growth patterns in {\em Arabidopsis} roots [Fig.~\ref{fig1}(iv),~\ref{fig1}(IV)]~\cite{kolb2017, migliaccio2001, migliaccio2013, thompson2004} and previous work on the packing of thin rods in granular media~\cite{schunter2018, algarra2018, mojdehi2016, schunter2019}, we consider the packing of an elongating slender loop, where we observe the same packing transitions. In this Letter, using a combination of experiments and scaling analysis, we characterize the emergent behavior of these distinct packing morphologies, and the role played by the evolution of the surrounding granular medium. These elastogranular systems will be helpful in the study of piercing \& penetration at soft-solid interfaces~\cite{choumet2012, cerkvenik2017}, in the design of dirigible surgical tools~\cite{swaney2012}, and provide a novel approach for looking at the packing of thin elastic structures across a spectrum of confinement strengths~\cite{chaudhuri2010, vetter2014, gosselin2014, schunter2019}.

Experimentally, a quasi-statically elongating elastic loop is confined to deform within granular monolayers at varying initial prepared packing fractions $\phi_0$~\footnote{Very similar deformations and packing transitions were observed when an elastic strip with a free boundary was inserted into a granular array, however the elastic loop removes any complications with how and where the free edge of the elastica finds the edge of the container.}. Soft, spherical hydrogel grains of radius $r=9.28\pm0.164$ mm comprise a granular bulk of approximately monodisperse, nearly frictionless grains (MagicWaterBeads). A long strip of polyethylene terephthalate (PET) film, identical to that used in reel-to-reel cinema projection, is clamped within a custom-built film sprocket/roller mount to form a ``pinched" elastic loop~\cite{domokos2003, santillan2005}, protruding into the experimental enclosure, of initial arclength $S_0 \approx 70$ mm, width $b=35$ mm (out of the page in Fig.~\ref{fig1}), and thickness $h=0.138$ mm [see videos S1 \& S2]. This mounting device creates a single clamped-roller boundary condition that allows for incremental adjustments $\Delta$ to the loop's arclength, up to a maximum value $S_0 + \Delta \simeq S_t/2$ (where the film's linear length $S_t=2.4352$ m). The geometry defined by the loop-tip ensures strictly elastic deformations of the material, as beginning an experiment with the film outside the enclosure would require plastic deformation.

\begin{figure}
\resizebox{1.0\columnwidth}{!}{\includegraphics{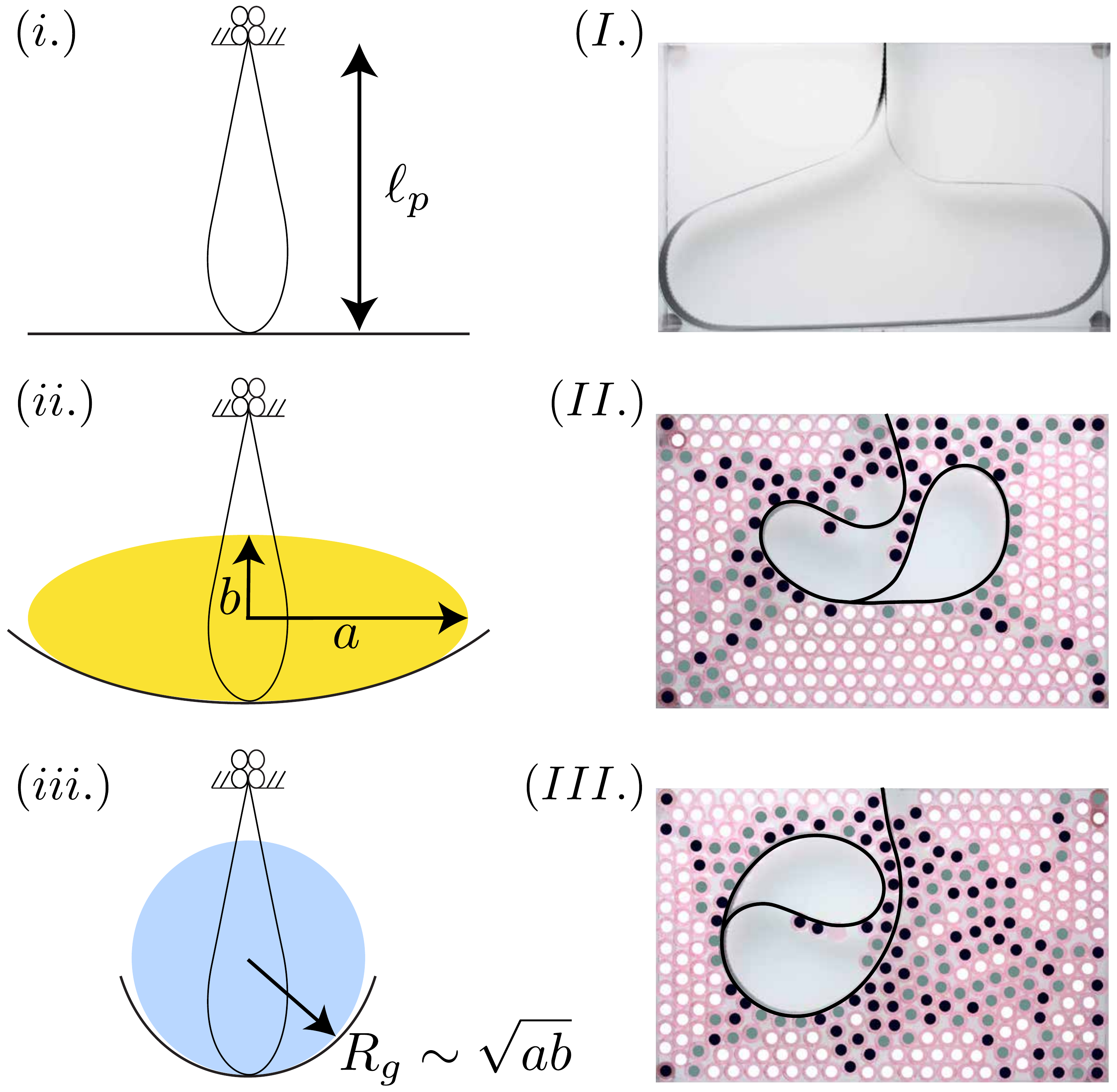}}
\caption[]{The role of the granular contour. While a freely injected elastic loop (i) will drape against a flat surface (I), the introduction of the granular medium can modify the curvature (ii,iii) and rigidity (II,III) of the surface against which the loop deforms. The influence of local bond orientation order $\psi^6$, which provides a measure of confining boundary rigidity, is apparent in both (II) folded and (III) circular packing morphologies. (\protect\tikz \protect\draw[blue,fill=white] (0,0) circle (.5ex);) $\psi^6 \geq 0.66$; (\protect\tikz \protect\draw[gray,fill=gray] (0,0) circle (.5ex);) $0.33<\psi^6<0.66$; (\protect\tikz \protect\draw[black,fill=black] (0,0) circle (.5ex);) $\psi^6 \leq 0.33$. The experiments shown in (I-III) are at the same injected arclength.}
\label{psi6} 
\end{figure}

As the arclength begins increasing by a small amount $\Delta$ ($\sim2$ mm), the elastic loop maintains its characteristic racket shape, a geometry observed over a wide range of length scales in fluid-thin structure interactions~\cite{py2007, cohen2003}. In the absence of any externally applied forces, the left-right symmetry of this configuration hypothetically persists in the limit $\Delta\ggg1$, however, the presence of the granular medium acts to confine and buckle the elastic loop. Thin structures favor bending over stretching as a deformation response to applied forces; indeed at larger $\Delta$-values, the symmetry of the pinched-loop configuration is lost as the thin structure relaxes stored curvature (housed primarily in the distal tip region) [Figs.~\ref{fig1}(ii),~\ref{fig1}(II)]. Continued increase of the arclength, along with local re-arrangements in the granular medium [Fig.~\ref{psi6}], elicit one of two distinct packing morphologies in the elastic loop: a disordered {\em folded} phase [Fig.~\ref{fig1}(a)] observable over the entire range of initial packing fractions, $0\leq\phi_0\leq0.828$, and an ordered {\em circular} phase [Fig.~\ref{fig1}(b)] emerging only in higher density arrays above a critical initial packing fraction $\phi_c=0.641$~\footnote{Our terminology is chosen to maintain consistency with previous works on the packing of flexible structures in both rigid and flexible confinement~\cite{donato2002, boue2006, stoop2008, vetter2014, pineirua2013, katzav2006, stoop2011, shaebani2017}.}. 

The implication here is that the granular medium (once jammed) acts as the confining mechanism, possessing both a certain level of rigidity and degree of curvature (ranging from flat to semi-circular) that will drive the system to adopt one morphology over the other. Looking at the local orientational order $\psi^6$ provides a visual (and also quantifiable) means of assessing the rigidity in a given array: highly-ordered regions ({\em i.e.} $\psi^6\geq0.66$) are observed to form the walls/surface of a granular containing space against which the elastic loop deforms [Fig.~\ref{psi6}]. Determining the extent to which the surrounding grains form a curved surface is more subtle and requires looking at how elastic deformations in the loop influence the formation of these boundaries.

\begin{figure}
\resizebox{1.0\columnwidth}{!}{\includegraphics{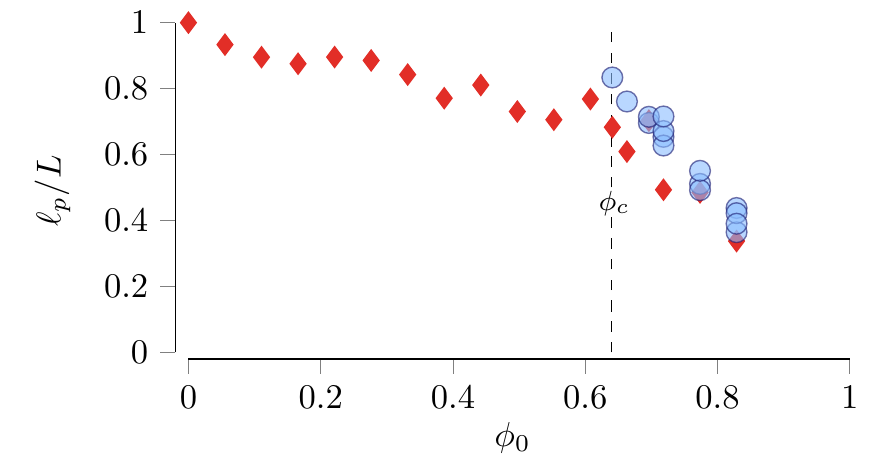}}
\caption[]{Penetration depth $\ell_p$ normalized by the length \textit{L} of the experimental enclosure, as a function of initial packing fraction $\phi_0$ for folding (red diamonds) and circular (light blue circles) geometries.}
\label{lp}
\end{figure}

We begin by quantifying the spatial extent of the elastic loop's deformation via a penetration depth $\ell_p$, the maximum distance (in the $y$-direction) which the thin structure can deform into an array at a given $\phi_0$, and a radius of gyration $R_g$, characterizing the (general) area within which elastic deformations localize. The measurement of penetration depth is straight-forward and progressively larger $\phi_0$-values are seen to result in lower values of $\ell_p$ overall, regardless of the thin structure's chosen morphology [Fig.~\ref{lp}]. The intuitive result that it becomes increasingly difficult to introduce additional arclength into a decreasing amount of available surface area, contrasts with the behavior that we observe above $\phi_c$, in which the circular morphology arises as a deformation mode. At equal $\phi_0$, penetration depths for circular packing are always greater than or equal to $\ell_p$-values measured in folded packing configurations [Fig.~\ref{lp}]. This behavior suggests that circular packing may be energetically preferable for thin structures elongating within dense granular media ($\phi_0>\phi_c$) in finite domains, commonly observed in root-bound plants in need of re-potting~\cite{amoroso2010}.

Values of $\ell_p$ remain nearly constant (at their maximum value) after the initial buckling of the loop, however, inklings of the final morphology only become apparent well into the post-buckling regime [Figs.~\ref{fig1}(ii),~\ref{fig1}(II)]: the loop continues to pack into the grains, which will eventually jam the surrounding granular network. We quantify the evolution of the surrounding medium towards a curved/circular profile by defining a radius of gyration $R_g$. Recall that the area of an ellipse (with semi-major/semi-minor axis' $\{a,b\}$) $A_e=\pi ab$, is equivalent to a circular area $A_c=\pi R_g^2$, whose radius we take as defining the radius of gyration $R_g$. This approach allows us to measure $R_g$ in all experiments, as the elongating loop tends to accumulate in elliptically-bounded regions below $\phi_c$ and in folded configurations [Fig.~\ref{psi6}(ii)]. Balancing terms between $\{A_e, A_c\}$ shows that the radius of gyration: $R_g \sim \sqrt{ab}$ (where for circular morphologies $a=b$) [Fig.~\ref{psi6}(iii)]. 

Measuring $R_g$ over the range of $\phi_0$-values [inset, Fig.~\ref{RgLc}(a)] (using the freely-available ImageJ platform~\cite{rasband2011}), we observe that for granular arrays prepared at low to mid-range packing fractions, $\phi_0 \lesssim 0.6$, the elastic loop exclusively adopts a folded geometry. In this regime the elastic loop is only weakly confined~\cite{vetter2014}. This behavior persists up until a critical packing fraction $\phi_c=0.641$, where we begin to see the emergence of the circular morphology. This does not imply that we always observe circular packing for $\phi_0 \geq \phi_c$, only that conditions within the system are now favorable for its emergence. The absence of the circular morphology at lower values of $\phi_0$ reinforces the argument made previously with $\psi^6$ that a certain strength of confinement ({\em i.e.} level of rigidity) is needed within the grains to observe circular packing. The grains must be able to form and maintain a semi-circular boundary contour against which the elastic loop deforms. 

\begin{figure}
\resizebox{1.0\columnwidth}{!}{\includegraphics{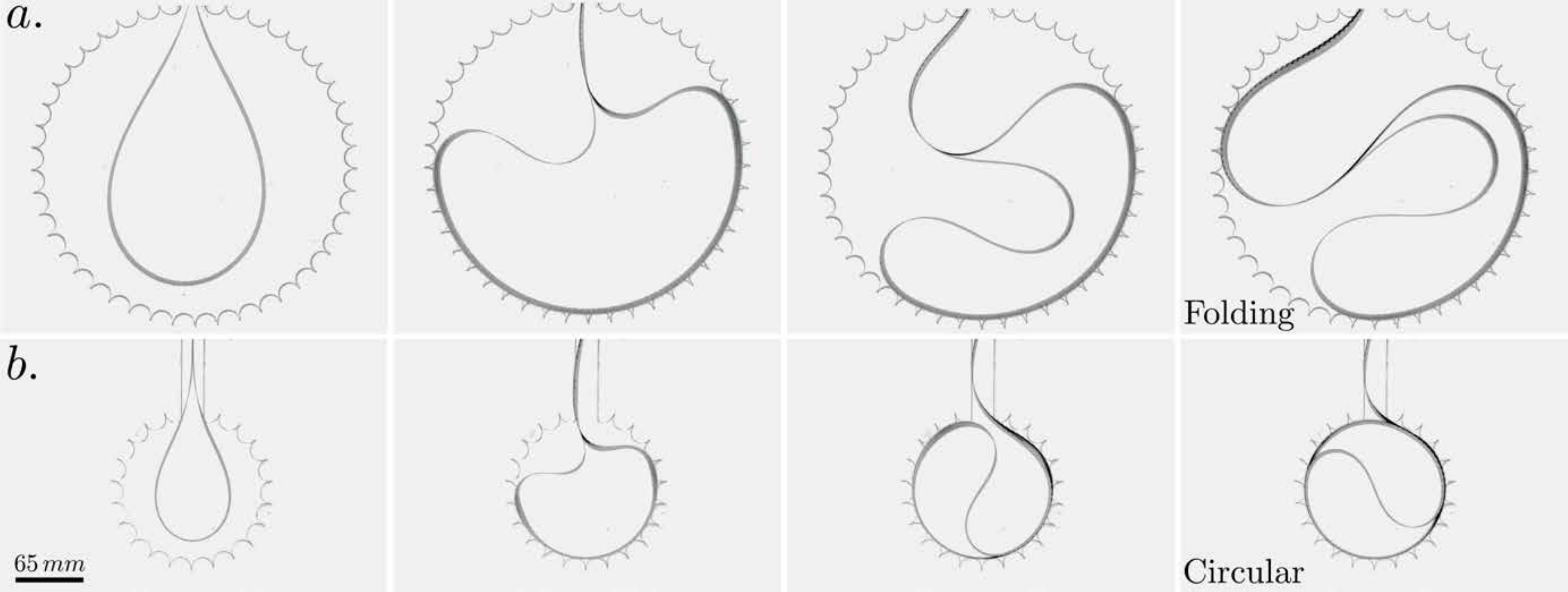}}
\caption[]{(a) Injection of an elastic loop into a rigid container with $R=117.78$ mm leads to periodic folding, while (b) injection into a container with $R=54.51$ mm leads to circular folding. The walls were cut with periodic rounded features to reduce the contact area, and thus the friction, between the loop and the wall.}
\label{figRigid}
\end{figure}

\begin{figure*}
\resizebox{1.0\columnwidth}{!}{\includegraphics{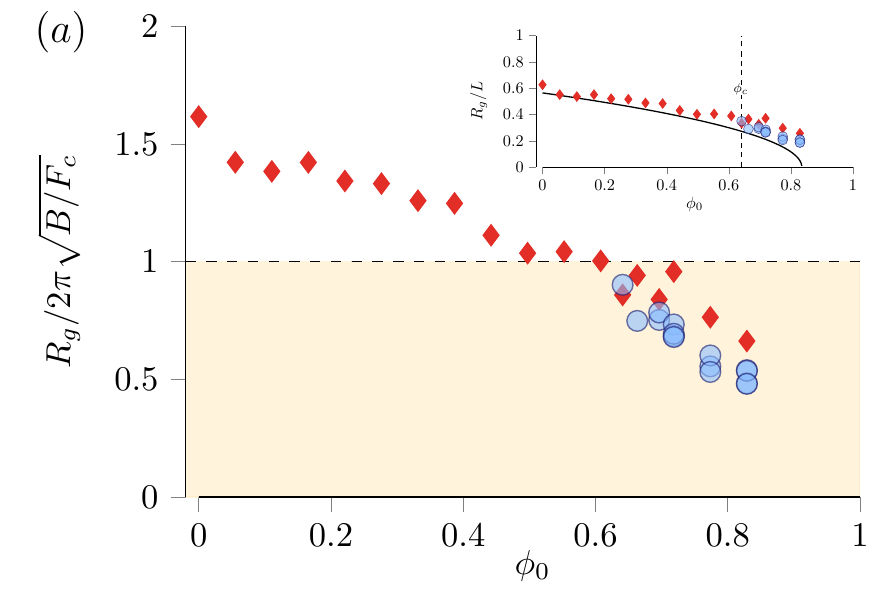}}\hspace{5mm}
\resizebox{1.0\columnwidth}{!}{\includegraphics{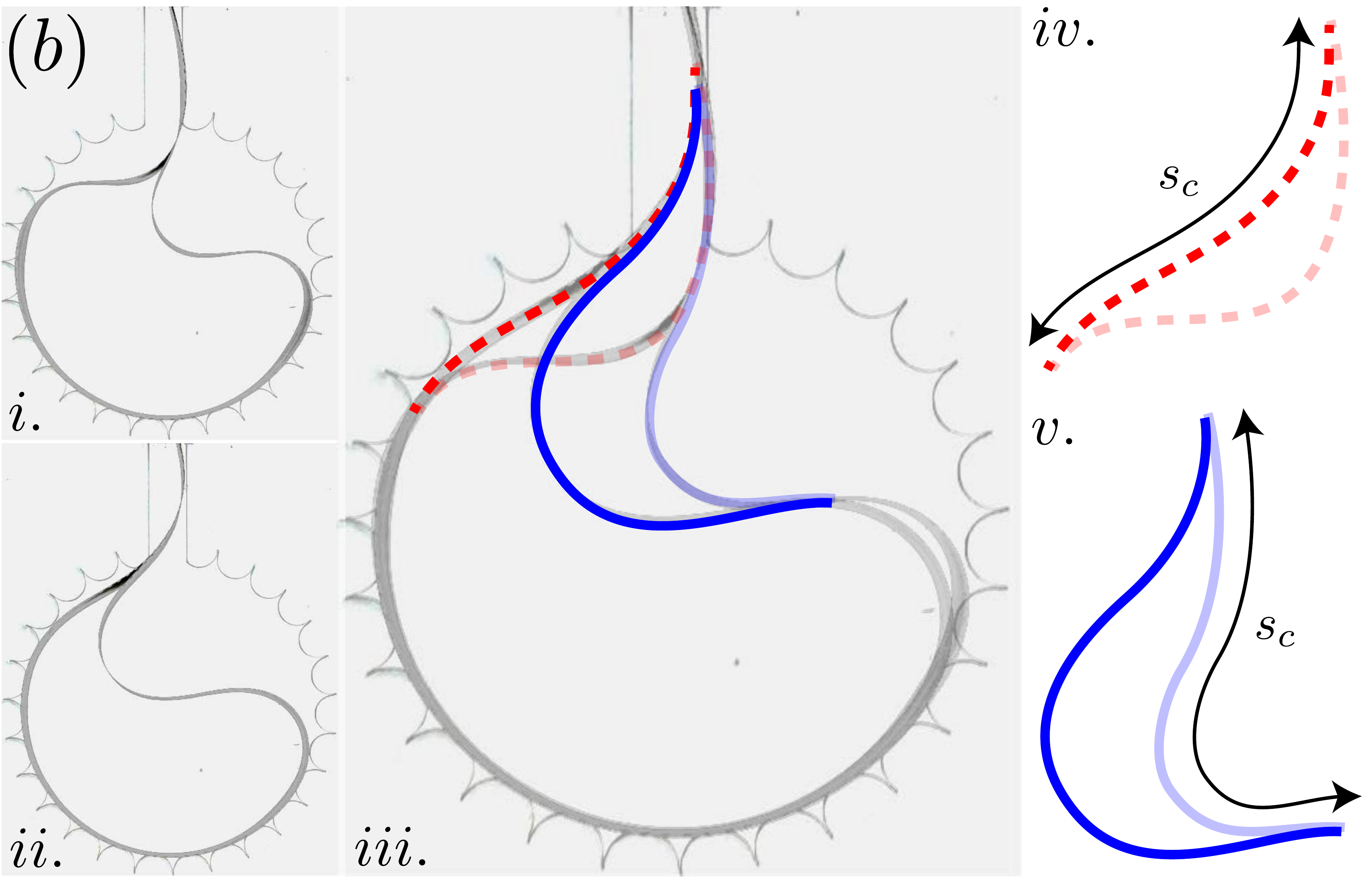}}\vspace{-2mm}
\caption[]{(a) Rescaling the radius of gyration $R_g$ by the critical length $L_c=2\pi\sqrt{B/F_c}$ brings material considerations into the problem, which are seen to be independent of system size. Notably, the circular packing morphology emerges when $R_g$ and $L_c$ are of the same order-of-magnitude. $\textbf{Inset:}$ Radius of gyration $R_g$, normalized by the length \textit{L} of the experimental enclosure, as a function of initial packing fraction $\phi_0$ for folding (red diamonds) and circular (light blue circles) geometries. (b) Images of the elastic loop $i.$ just prior to buckling and $ii.$ immediately following buckling. $iii.$ Images from $i.$ and $ii.$ overlaid on each other, with the two regions of the loop that appear to buckle highlighted (dashed red; solid blue). The length of the $iv.$ left and $v.$ right half of the loop that appears to buckle are labeled $s_c$. The labeled arclengths are both longer than $R_g$, meaning this structure will pack with a circular morphology.}
\label{RgLc}
\end{figure*}

We speculate that the confining geometry necessary for the transition between folded and circular patterns is defined by a critical radius of gyration $R_c$. Given a 2D array of grains with packing fraction $\phi_0 \geq \phi_c$, $R_c$ is the circular inclusion that would cause the array to jam locally, forming an effectively rigid containing region within which the slender loop will pack. It was previously found~\cite{schunter2018} that the soft hydrogel grains used in these experiments become jammed at a critical packing fraction $\phi_j=0.8305\pm0.0135$. As $\Delta$ increases in the limit that $\phi \to \phi_j$, the total surface area (available to the grains) within the array will change by an amount proportional to a circular area $\sim \pi R_c^2$, such that: $\phi_j \sim N \pi r^2/(LW-\pi R_c^2)$~\footnote{We can account for the surface area of the film by either ignoring it in the limit $h\lll1$, a safe assumption here, or considering it as being included in the measurement of $R_g$.}. Rearranging to isolate for $R_c$ yields the scaling for the critical radius of gyration as:
\begin{equation} \label{1}
R_c \sim \sqrt{\left(\frac{LW}{\pi}\right)-\left(\frac{Nr^2}{\phi_j}\right)}\,.
\end{equation}
The RHS of Eq.$\,(\textbf{\ref{1}})$ is composed entirely of known values, which yields $R_{c} \simeq 94.32$ mm. We corroborate this scaling argument by performing a set of additional ``toy model" experiments where the elastic loop is injected into circular, rigid walled confinements of different sized internal diameters. This has the effect of removing the granular part of the problem and creates a truly rigid confining boundary with which we can test our initial assumption of treating the jammed grains as a rigid, effectively curved surface above $\phi_c$ [Fig.~\ref{figRigid}; videos S3 \& S4]. Experiments using this toy model produce a value for the critical radius of gyration $R_{t} \simeq 99.12$ mm, close to the value obtained using the scaling in Eq.$\,(\textbf{\ref{1}})$. 


Although the grains are absent in the toy model experiments, we can still infer hypothetical values of $\phi_0$ associated with each $R_g$ tested for these idealized rigid boundaries. A simple rewriting of Eq.$\,(\textbf{\ref{1}})$ yields the hypothetical $\phi_0$-value for a particular rigid circular confinement region found in the toy model experiments [Fig.~\ref{figRigid}] as:
\begin{equation} \label{2}
\phi_0 \sim \phi_j\left(1-\frac{\pi R_g^2}{LW}\right)\,.
\end{equation}
Rearranging to isolate for $R_g/L$, Eq.$\,(\textbf{\ref{2}})$ is plotted as a solid black line in the inset of Fig.~\ref{RgLc}(a), alongside experimental measurements for both folded (red diamonds) and circular (light blue circles) runs. The scaling $(\textbf{\ref{2}})$ is in very good agreement with experiments; it validates $R_g$ as a measure of system physics and provides a lower bound to the experimental data expressing the provisional/nominal system size~\cite{vetter2014}. 

Normalizing $R_g$ by an arbitrary system dimension [inset, Fig.~\ref{RgLc}(a)] creates {\em de facto} system-size dependence. Ideally, we want to be able to describe these elastogranular interactions in a scale-invariant way. As a length scale of granular origin, $R_g$ can be regarded as a proxy for the geometric constraints in a given array. Properly accounting for global shape changes, specifically the coupling between large deformations of the thin loop structure and the evolution of its granular containing space, requires that we bring material considerations into the analysis. Drawing on similarities with other physical systems utilizing a thin loop~\cite{wang1981, cohen2003, mora2012, sano2017, bico2018} hints at the existence of an additional length scale originating with the slender structure.

We assume that an effectively rigid, locally jammed region of the granular boundary generates a force proportional to the reaction force measured at the jamming point $\phi_j$ for these particular grains~\footnote{The jamming point $\phi_j$, calculated in separate experiments~\cite{schunter2018}, coincides with a sudden and rapid increase of the reaction force measured in continuous, quasi-statically compressed granular arrays~\cite{schunter2019}. We take this critical reaction force in the granular array at jamming as $F_c$.}. Prior to buckling, forces between the grains and loop tip will balance; the bending rigidity $B=EI$ of the elastic loop (calculated using the ``free-fold test''~\cite{plaut2015}) opposes any breaking of symmetry. Treated as an Euler-buckling problem, there exists a critical load $F_c =4\pi^2 B/L^2$, above which the loop will buckle. Balancing the critical reaction force $F_c$ in the granular array at jamming with the material properties of the elastic loop, we arrive at the critical length
\begin{equation} \label{Lc}
L_c=2\pi\sqrt{B/F_c}. 
\end{equation}
Fig.~\ref{RgLc}(a) shows $R_g$ as a function of $\phi_0$, where $R_g$ has been rescaled by the critical length $L_c$. Circular packing necessitates buckling of the slender loop (allowing it to accommodate excess length or ``buffer-by-buckling"~\cite{vella2019}) and the presence of a circular confining boundary, conditions which become possible when $R_g$ and $L_c$ are of equal orders-of-magnitude: 
\begin{equation} \label{3}
R_g = R_c \sim L_c\,.
\end{equation}
The existence of this length scale helps illuminate the lack of physical information provided by the penetration depth: $L_c$ is the length scale of elastic buckling. While we calculate $L_c$ from independent measurements of $B$ and $F_c$, a natural question is: what length of the elastic loop does $L_c$ correspond to? Returning to the toy model experiments, we isolate two images of the elastic loop just prior to buckling [Fig.~\ref{RgLc}(b)--i.] and just after buckling [Fig.~\ref{RgLc}(b)--ii.] Overlaying these images [Fig.~\ref{RgLc}(b)--iii.], we identify two portions of the elastic loop that appear to buckle at the same time (one on the left half of the loop, the other on the right half). Isolating each length that appears to buckle, and labeling it $s_c$, we would expect these structures to pack with a circular morphology if we take $L_c \equiv s_c$ and find that $s_c > R_g$. We find that both lengths are longer than the radius of gyration containing the loop ($R_g=67.2$ mm; $s_c=133.8$ mm [dashed red], $s_c=141.2$ mm [solid blue]), which suggests these lengths that appear to buckle are what drives the selection of the packing morphology~\footnote{The transition between periodic folding and circular packing occurs when $R_g/L_c < 1$. However, the curvature near the elastic loop is dictated by the local arrangement of grains. If the grains pack with a large radius of curvature, the loop will pack periodically, even if $R_g/L_c<1$. To confirm this using our toy model, we injected an elastic loop into an ellipse of $R_g=67.2$ mm [see video S5], and find that it folds periodically even though packing into a circular profile with equal $R_g=67.2$ mm led to circular packing [Fig.~\ref{figRigid}; video S4].}.

\begin{figure}
\resizebox{1.0\columnwidth}{!}{\includegraphics{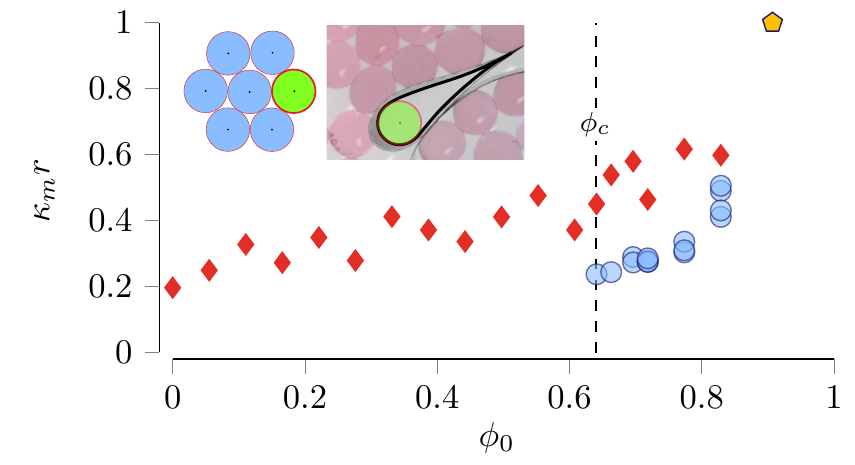}}
\caption[]{Normalized maximum measured curvature $\kappa_m$ as a function of initial packing fraction $\phi_0$ for folding (red diamonds) and circular (light blue circles) geometries. The yellow pentagon represents the anticipated limiting behavior of this elastogranular system $\textbf{Inset:}$ As $\phi_0 \to 0.907$, the maximum observable loop-tip curvature will approach a limiting value $\kappa_{h}$, with a radius of curvature on the order of the average size grain radius in an array.}
\label{kappa}
\end{figure}

In circular morphologies, the energy minimization strategy for the elastic loop is essentially fixed once the circular profile is formed. Additional injected arclength will wrap around this inner circle similar to DNA spooling within a capsid~\cite{phillips2012}. We can measure this radius value experimentally, or infer a value using the scaling in $(\textbf{\ref{1}})$. In folded morphologies, we observe the formation of a cascade of loops increasing in number as either $\Delta$ or $\phi_0$ become larger, with loop-tip curvature values appearing to approach a limiting value $\kappa_{h}$. A similar situation has been observed numerically in elastic rings with self contact~\cite{flaherty1972}. The limiting curvature $\kappa_{h}$ is the largest amount the loop may be bent by the granular medium without inducing any plastic deformation. Due to the monodispersity of the grains, we know {\em a priori} the granular medium will approach a hexagonal-packing configuration as an equilibrium geometry~\cite{vanhecke2009, curk2019}. We thus anticipate that a loop-tip has a limiting radius of curvature, ${1/\kappa_h}$, on the order of the average size grain radius $r$ [inset, Fig.~\ref{kappa}]. 

Along with material considerations, how slender structures deform depends intimately on their boundary interactions. For the canonical case of elongating thin objects confined within rigid containers, the geometry, stiffness, and boundary continuity of the confining space can each contribute to the deformation morphologies adopted by these objects. By studying elastogranular packing, we are able to probe a system in which large elastic deformations occur within transitional boundaries. The discrete, flexible network of the granular medium surrounding the loop (at low packing fractions) will change under increasing confinement, becoming more analogous to a continuous, rigid containing space. Simple experimental systems such as these provide a novel investigative tool for looking at how macroscale geometric features can arise in complex media, and how these features in turn affect interactions with inclusions. 


\section*{Conflicts of interest}
There are no conflicts to declare.

\section*{Acknowledgements}
The authors wish to thank Rinik Kumar for performing preliminary experiments on the free-boundary elastica. We are grateful for financial support from the National Science Foundation CMMI--CAREER through Mechanics of Materials and Structures (No. 1454153).



\bibliography{eGL} 
\end{document}